\documentclass[pra,aps,amsmath,nofootinbib,superscripataddress,showpacs]{revtex4}
\usepackage{graphicx}
\usepackage{hyperref}
\usepackage{bm}
\usepackage{epsfig}
\usepackage{amsfonts}
\usepackage{mathrsfs}
\usepackage{amsmath}
\catcode`@=11
\let\savesort=\NAT@sort@cites
\newcommand\nosort[1]{\edef\NAT@cite@list{#1}}
\def\citenosort#1{\let\NAT@sort@cites=\nosort\cite{#1}%
 \let\NAT@sort@cites=\savesort}
 \catcode`@=12% at signs are no longer letters
\makeatletter
  
  \newcommand{\Rmnum}[1]{\expandafter\@slowromancap\romannumeral #1}
\makeatother

\newcommand{\beq}{\begin{equation}}
\newcommand{\eeq}{\end{equation}}
\newcommand{\bea}{\begin{eqnarray}}
\newcommand{\eea}{\end{eqnarray}}
\newcommand{\pa}{\partial}
\newcommand{\bib}{\bibitem}
\newcommand{\comment}[1]{}
\newcommand{\ep}{\varepsilon}
\newcommand{\kR}{\kappa_R}
\newcommand{\la}{\lambda}
\newcommand{\kRt}{\tilde{\kappa}_R}

\def\mO{{\mathcal O}}

\begin{document}

%%%%%%%%%%%%%%%%%%%%%%%%%%%%%%%%%%%%%%%%%%%%%%%%%%%%%%%%%%%%%%%%%%%%%%%%%%%%%%%%%%%%%%%%%%%%%%%
%%%%%%%%%%%%%%%%%%%Define Title, Author, Address, Preprint#
\title{Conditions for supersonic bent Marshak waves}
\author{Qiang Xu}
\email{xuqiangxu@pku.edu.cn}
\affiliation{Key Laboratory of Pulsed Power, Institute of Fluid Physics, CAEP, P. O. Box 919-108, Mianyang 621999, China}
\author{Xiao-dong Ren}
\affiliation{Key Laboratory of Pulsed Power, Institute of Fluid Physics, CAEP, P. O. Box 919-108, Mianyang 621999, China}
\author{Jing Li}
\affiliation{Key Laboratory of Pulsed Power, Institute of Fluid Physics, CAEP, P. O. Box 919-108, Mianyang 621999, China}
\author{Jia-kun Dan}
\affiliation{Key Laboratory of Pulsed Power, Institute of Fluid Physics, CAEP, P. O. Box 919-108, Mianyang 621999, China}
\author{Kun-lun Wang}
\affiliation{Key Laboratory of Pulsed Power, Institute of Fluid Physics, CAEP, P. O. Box 919-108, Mianyang 621999, China}
\author{Shao-tong Zhou}
\affiliation{Key Laboratory of Pulsed Power, Institute of Fluid Physics, CAEP, P. O. Box 919-108, Mianyang 621999, China}
\begin{abstract}
Supersonic radiation diffusion approximation is a useful way to study the radiation transportation. Considering the bent Marshak wave theory in 2-dimensions, and an invariable source temperature, we get the supersonic radiation diffusion conditions which are about the Mach number $M>8(1+\sqrt{\ep})/3$, and the optical depth $\tau>1$. A large Mach number requires a high temperature, while a large optical depth requires a low temperature. Only when the source temperature is in a proper region these conditions can be satisfied. Assuming the material opacity and the specific internal energy depend on the temperature and the density as a form of power law, for a given density, these conditions correspond to a region about source temperature and the length of the sample. This supersonic diffusion region involves both lower and upper limit of source temperature, while that in 1-dimension only gives a lower limit. Taking $\rm SiO_2$ and the Au for example, we show the supersonic region numerically.
\end{abstract}

\noindent

\maketitle

\section{Introduction}
\label{sec:introduction}
Radiation transport in plasma is complicated because of complicated course between photon and material. The local material density and temperature affect the emission
and absorption of the radiation, while the non-local radiation affects the radiation energy flux. When the material is optically thin, the radiation flux can bleach through and ionize the
material, but when the material is optically thick, the photons of the external source are absorbed and re-emitted many times before reaching the heating front. When the opacity
is so large that the mean free path of photon is much less than the distance over which temperature are changing, the radiation diffusion approximation can be used.
Assuming the radiation is isotropic, and considering the gray diffusion approximation, the radiation diffusion act as heat waves just like the thermal conduction. Radiation heat waves as a specifical form of radiation diffusion are very common and important in the plasma physics, and in particular in inertial confinement fusion(ICF) where we need to know how the laser energy
is absorbed by the hohlraum inner wall. It also appears in the astrophysical situations such as supernova explosions and the penetration of radiation into the interstellar medium.

Generally speaking, the radiation diffusion can be described as the interaction between radiation and material. Most of the time, researchers tend to study the supersonic radiation diffusion in which the time scales of radiant
energy exchange are quite smaller than hydrodynamic time scales, so that the penetration of radiation and energy deposition can occur while the density of material does not change yet. That's to say, we can treat the density of material as a constant in supersonic radiation diffusion. This treatment in supersonic radiation diffusion condition brings significant convenience to both the theory and the experiment.

The earliest solution of the supersonic heat waves was worked out in 1958, by Marshak\cite{Marshak} who demonstrated that the location of radiation front is proportional to $\sqrt{t}$, and after the front, the radiant flux is taken up by the material and converted into internal energy. Following the work of Marshak, many more accurate solutions are worked out\cite{Ya,Mihalas,Castor,Smith,Hammer}. However, these solutions are entirely confined into one-dimensions which is difficult to describe the bent shape of the heat front. In 2000, using $\Omega$-laser, Back et al designed an experiment to study the supersonic radiation diffusion in (50mg/cc)$\rm{SiO_2}$ and (40mg/cc)$\rm{Ta_2O_5}$ and found the curvature of the radiation front\cite{Back}. To explain such a curvature, Hurricane and Hammer developed an analytic solution(Bent Marshak wave) in two-dimensions\cite{Hammer1}. This theory indicates that energy loss to the sleeve which surrounds the material causes the curvature of the radiation front.

Experimentally, in addition to the Back's experiments, Afshar-ra and Hoarty et al studied the supersonic heat wave in low density foam composed of CH compound and a little other elements used for determining temperature \cite{rad,Hoarty,Hoarty1,Hoarty2,Massen,Peterson}, and Bozier did the similar work in the high Z gas composed of Xe\cite{Bozier}. Generally, when designing the supersonic radiation diffusion experiment, one must consider two conditions. first, the speed of the radiation front must be larger than the sound speed of the material which requires a high temperature and a low density of the material. Second, to use the radiation diffusion approximation, the material must be optically thick that requires a low temperature and a high density. The two conditions have a opposite dependence on the temperature and density of material. Therefore, the temperature and the density must satisfy a specifical condition so that the supersonic radiation diffusion approximation can be used. On the other hand, when this radiation heat wave propagates in the material, the temperature drops, so does the wave's speed. So, how long the supersonic radiation heat wave can propagate is a problem which need to be worked out.

In one-dimension, the condition for the supersonic radiation has been shown in \cite{Wu}. In this paper, considering a power-law opacity, we discuss the condition for the supersonic bent marshak wave in two-dimensions, and study how long the supersonic Marshak wave can propagate. This paper is organized as follows. In Sec. I, we introduce the analytic solution of bent Marshak wave. In Sec. II, we generally discuss the condition for supersonic Marshak wave. In section III, we study the  condition for supersonic Marshak wave theoretically by using $\rm{SiO_2}$ and Au for example. In sec IV, we give the numerical result. At last, we give the conclusions in Sec. IV.

\begin{figure}
\psfig{figure=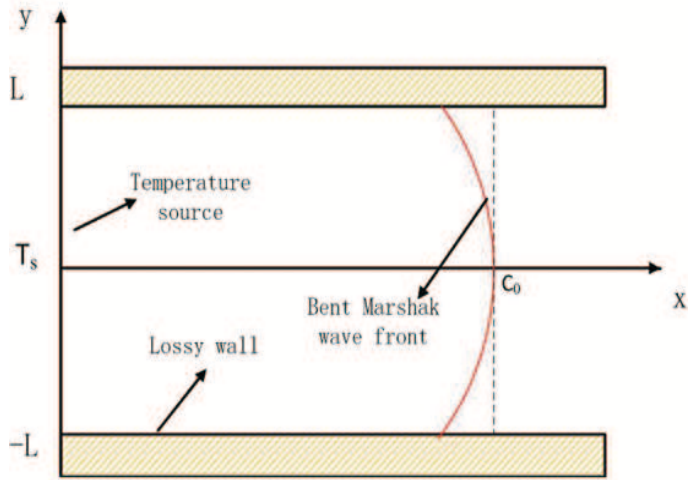,width=9cm,height=7cm}
\caption{The picture that bent heat waves propagate in a pipe with a constant temperature source on the left and a lossy wall
bounds the problem at y= ¡ÀL.}
\label{bent}
\end{figure}

\section{Basic theory of Bent Marshak wave}
\label{bto}
\subsection{The expansion of the solution and boundary conditions}
The propagating of heat waves can be described as the interaction between radiation and the material. In this course, the material gets internal energy from the radiation flux, which can be expressed as\cite{dovich}
\beq
\label{trans}
\rho\frac{\pa e}{\pa t}=\frac{4}{3}\nabla\left[\frac{1}{\rho\kappa_R}\nabla(\sigma T^4)\right],
\eeq
where where $\rho$ is the material density, $e$ is the material specific
internal energy, T is the material and
radiation temperature when considering the local thermodynamic equilibrium, $\rm\sigma=1.03\times10^{12} ergs/ (cm^2¡¤s¡¤eV^4)$ is
the Stefan-Boltzman constant, and $\kappa_R$ is the Rosseland mean opacity.

In the supersonic case, considering the density $\rho$ and Rosseland mean opacity as a constant, and away from the radiation front, the general solution of $T^4(x,t)$ in two-dimensions pipe as FIG. \ref{bent} can be given as

 \beq
\label{gs}
T^4=\sum_{n=0}^{\infty}\cos(k_ny)[A_n(t)e^{k_nx}+B_n(t)e^{-k_nx}],
\eeq
where $k_n$ is decided by the boundary conditions, $A_n$ and $B_n$ are time-dependent coefficients respectively. According to the form of temperature, the form of the radiation front can be chosen to be

 \beq
\label{front}
x_f(x,t)=\sum_{n=0}^{\infty}c_n(t)cos(k_ny),
\eeq
which describes a bent radiation front.
Hurricanea and Hammer think that the bent shape of the radiation front is caused by a non-ideal
boundary at the walls, $y=\pm L$, which has a albedo, $a$. Not all the energy flux into the boundary can be reflected as a form of re-emission, so, the boundary condition at walls is defined as\cite{Hammer1}

\beq
\label{boundary}
F|_{y=\pm L}=-\frac{4}{3\rho\kappa_R}\nabla(\sigma T^4)=-(a-1)\sigma T^4,
\eeq

while the boundary condition at two sides are $T|_{x=0}=T_s$ and $T|_{x=x_f}=0$ at  respectively.
According to the boundary conditions at walls and the general solution of temperature, ones get the eigenvalue condition

\beq
\label{eig}
{\rm tan}(k_nL)=\frac{\ep}{k_nL},
\eeq
which gives the eigenvalues
\bea
\label{kn}
\begin{aligned}
k_0&=\frac{\sqrt{\ep}}{L},\\
k_n&=\frac{n\pi}{L}+\frac{\ep}{Ln\pi}, n\geq1,
\end{aligned}
\eea

where $\ep=\frac{3}{4}\rho\kappa_RL(1-a)$ is a small value, considering the albedo of walls tends to be 1. Hurricanea and Hammer expand the solution, Eq. (\ref{gs}) in $\sqrt{\ep}$ and give\cite{Hammer1}

\bea
\label{kn}
\begin{aligned}
\frac{T^4}{T_s^4}&=-(1+\frac{\ep}{3}){\rm cos}(\sqrt{\ep}\frac{y}{L})\frac{\sinh[k_0(x-c_0)]}{\sinh(k_0c_0)}\\
&+\frac{4\ep}{\pi^2}\sum_{n=1}^{\infty}{\rm cos}(k_ny)\frac{(-1)^{n+1}}{n^2}\frac{\sinh[k_n(x-c_0)]}{\sinh(k_nc_0)}+\mO(\ep^{3/2}),
\end{aligned}
\eea
where $c_0$ is zeroth coefficient of the radiation front, and we can approximately consider it as the position of the front for leading order.
\subsection{Bent radiation front}
According to the expression of temperature, combining with the equation of motion(EOM) of the radiation front\cite{Hammer}

\beq
\label{EOM}
\rho e\dot{x}_f=\left.-\frac{4\sigma}{3\kR\rho}\nabla T^4\right|_{x_f},
\eeq
found by Hammer and Rosen, where $e$ is the specific internal energy of material at $T_s$, the coefficients of the radiation wave front can be restricted as\cite{Hammer1}

\bea
\label{cn}
\begin{aligned}
\dot{c}_0&=\frac{4\sigma T_s^4}{3\rho^2\kR e}(1+\frac{\ep}{3})\frac{\sqrt{\ep}}{L\sinh(\sqrt{\ep}c_0/L)}+\mO(\ep^2),\\
\dot{c}_n&=\frac{8\sigma T_s^4}{3\rho^2\kR e}\frac{(-1)^n\ep}{n^2\pi^2}\frac{k_n}{\sinh(k_nc_0)}+\mO(\ep^2), n\geq1.
\end{aligned}
\eea
Considering the leading order, thinking $c_0\simeq x_f$ and integrating Eq. (\ref{cn}), the expression of the front is given as
\bea
\label{cn1}
\begin{aligned}
x_f(y,t)\simeq\frac{L}{\sqrt{\ep}}{\rm cosh}^{-1}\left[\frac{D\ep t}{2L^2}+1\right]{\rm cos}(\sqrt{\ep}y/L),
\end{aligned}
\eea
where the parameter $D=(1+\ep/3)\frac{8\sigma T_s^4}{3\rho^2\kR e}$ which represents the velocity of the radiation front is just like the diffusion constant of the Marshak wave\cite{Marshak}. The parameter $\ep$ decided
by the albedo, determines the shape of the radiation front. According to Eq. (\ref{cn1}), if the wall absorb more energy and re-emission less energy, the parameter $\ep$ will be larger, and the curvature of the radiation front will also be larger.

\section{Conditions for radiation-dominated and supersonic }
\label{cfr}
Radiative energy density is small compared
to the material energy density, therefore it often does not
dominate the energy balance. However, if the speed of radiation front is much greater than the sound speed, the radiative energy flux dominates the material energy flux and therefore can determines energy flow. In some way, we think the conditions for Radiant energy flux domination and the supersonic are equal approximately. Generally, the radio of these two term,

\bea
\label{radio}
{\rm\gamma=\frac{radiative ~energy~ flux}{material~ energy ~flux}=\frac{\sigma T^4}{\rho eC_S}},
\eea
must be greater than 1. Setting $y=0$, we expand the form of the radiation front and get

\bea
\label{xf1}
\left.x_f\right|_{y=0}\approx\sqrt{Dt}-\frac{L}{12}\sqrt{\frac{2}{\ep}}\left(\frac{D\ep t}{2L^2}\right)^{3/2}+\mO(\ep^{3/2}).
\eea
Using Eq. (\ref{xf1}), considering the velocity of radiation front $u$ is equal to $x_f/t$ approximately, and defining the optical depth $\tau\approx\rho\kR x_f=x_f/l_0$ ($l_0$ represents the mean free path of the photon), we get
\bea
\label{xf2}
\sigma T^4=u\tau\rho e\frac{3}{8(1+\ep/3)}.
\eea
Substituting Eq. (\ref{xf2}) into Eq. (\ref{radio}), we get
\bea
\label{radio1}
{\rm\gamma=\frac{3}{8(1+\ep/3)}\cdot M\cdot\tau},
\eea
where The Mach number, M, is the ratio of the radiative
front velocity to the heated material sound speed and its value increases as the importance of the radiative flux increases.

According to the definition, the radiative energy flux dominating requires
\bea
\label{radio2}
M\cdot\tau>\frac{8(1+\ep/3)}{3},
\eea
which is similar to the condition gotten by Back\cite{Back}. To satisfy the condition of diffusion approximation, we need $\tau=x_f/l_0>1$ which means the represents the mean free path of the photon is much less than the distance over which temperature is changing. On the other hand, we hope the hydrodynamic density perturbations
at the radiation front remain as small as possible. Since the changes of material density have some relationships with the velocity of the wave front, that is $\frac{\rho_2}{\rho_1}\approx(1+M^{-2})$\cite{Paul}.
The Mach number is required to be large enough that ones may ignore the density perturbation. Therefore, the conditions for radiation-dominated are
\bea
\label{cn2}
\begin{aligned}
M&>\frac{8(1+\ep/3)}{3}, ~\tau>1,
\end{aligned}
\eea
on which the hydrodynamic density perturbations
will be less than $10\%$.

We assume the opacity and the specific internal energy depend on temperature and the density as a form of power law\cite{Hammer,57,58,59}, which can be expressed as
\bea
\label{opacity}
\begin{aligned}
\kR&=\tilde{\kappa}_R\left(\frac{T_s}{T_0}\right)^{-\alpha}\left(\frac{\rho}{\rho_0}\right)^{\lambda}, \\
e&=e_0\left(\frac{T_s}{T_0}\right)^{\beta}\left(\frac{\rho}{\rho_0}\right)^{-\mu},
\end{aligned}
\eea
where $\alpha,~\la,~\beta,~\mu$ are the positive constant, $T_0$ and $r_0$ represent the specifical temperature and density respectively. At the same time, we can give the expression of sound
speed as
\bea
\label{Cs}
\begin{aligned}
C_s&=\sqrt{\frac{\pa P_i}{\pa\rho}+\frac{\pa P_e}{\pa\rho}}
=\sqrt{(1+Z)\frac{k_BT}{Am_p}}\\
&=\Gamma \left(\frac{T_s}{T_0}\right)^{1/2},
\end{aligned}
\eea
where $k_B$ and $m_p$ are Boltsmann constant and the mass of proton respectively, $\Gamma$ is a constant with a dimension $m/s$. Therefore the conditions of the radiation-dominated and supersonic
can be rewritten as
\bea
\begin{aligned}
M=\frac{\dot{x}_f}{C_s}=\frac{4\sigma T^4_0}{3\kRt \rho_0^2e_0\Gamma L}\left(\frac{T_s}{T_0}\right)^{4+\alpha-\beta}\left(\frac{\rho}{\rho_0}\right)^{\mu-2-\lambda}\\
(1+\ep/3)\frac{\sqrt{\ep}}{\sinh(\sqrt{\ep}x_f/L)}>\frac{8(1+\ep/3)}{3},\\
\tau=\rho\kR x_f=\rho_0\kRt x_f\left(\frac{T_s}{T_0}\right)^{-\alpha}\left(\frac{\rho}{\rho_0}\right)^{1+\la}>1,
\end{aligned}\label{cn4}
\eea
from which, we know a higher source temperature and the lower density lead to a longer distance that the supersonic heat wave can propagate noticing that the Mach number is proportional to $\sqrt{\ep}/\sinh(\sqrt{\ep}x_f/L)$.
\section{Numerical result for two kinds of material}
\label{nsf}
In the supersonic diffusion, the material density can be considered as a constant, and the optical depth is proportional to $x_f$. When radiation wave front reach a a critical position $x_1$, the optical depth become large enough that the diffusion approximation can be used. While the radiation heat wave is propagating in the material, the temperature becomes lower, and the velocity of the radiation front becomes smaller. When radiation front reach another critical position $x_2$, the velocity of the radiation wave front become subsonic. Therefore, when giving a certain density, every source temperature corresponds to a critical position. usually, $x_1$is a little smaller than $x_2$. That to say, for a given temperature, when the front position is in the region $x_1<x<x_2$, we can use the supersonic radiation diffusion approximation. We use two kinds of material, ${\rm SiO_2}$ and Au for example to illustrate this region.
\subsection{supersonic diffusion region for ${\rm SiO_2}$}
Taking ${\rm SiO_2}$ for example, considering it as ideal gas, and assuming the $\rm SiO_2$ plasma with a density $\rho_0=50mg/cc$ is fully ionized at the characteristic temperature $T_0=10^6k$, according to Eq. (\ref{Cs}), sound speed can be given as
\bea
\label{Cs1}
Cs(T)=6.74\times10^4[m/s]\left(\frac{T}{T_0}\right)^{1/2}.
\eea
The specific internal energy of ${\rm SiO_2}$ can be expressed as
\bea
\label{e0}
e&=\frac{k_BT}{(\gamma-1)Am_p}
&=6800[MJ/kg]\left(\frac{T}{T_0}\right).
\eea
By fitting the numerical result of opacity of ${\rm SiO_2}$, the paper \cite{Wu} give the relationship
\bea
\label{kRsio2}
\kR=175m^2/kg\left(\frac{T}{T_0}\right)^{-3.3}\left(\frac{\rho}{\rho_0}\right)^{0.64}
\eea
Therefore, the conditions for supersonic diffusion in ${\rm SiO_2}$ are given as
\bea
\begin{aligned}
0.474\left(\frac{T_s}{10^6k}\right)^{5.8}\left(\frac{\rho}{50mg/cc}\right)^{-2.64}\frac{\sqrt{\ep}}{\sinh(\sqrt{\ep}x_f/L)}>\frac{8}{3},\\
8.75\frac{x_f}{mm}\left(\frac{T_s}{10^6k}\right)^{-3.3}\left(\frac{\rho}{50mg/cc}\right)^{1.64}>1
\end{aligned}\label{cnsio3}
\eea

\begin{figure}
\psfig{figure=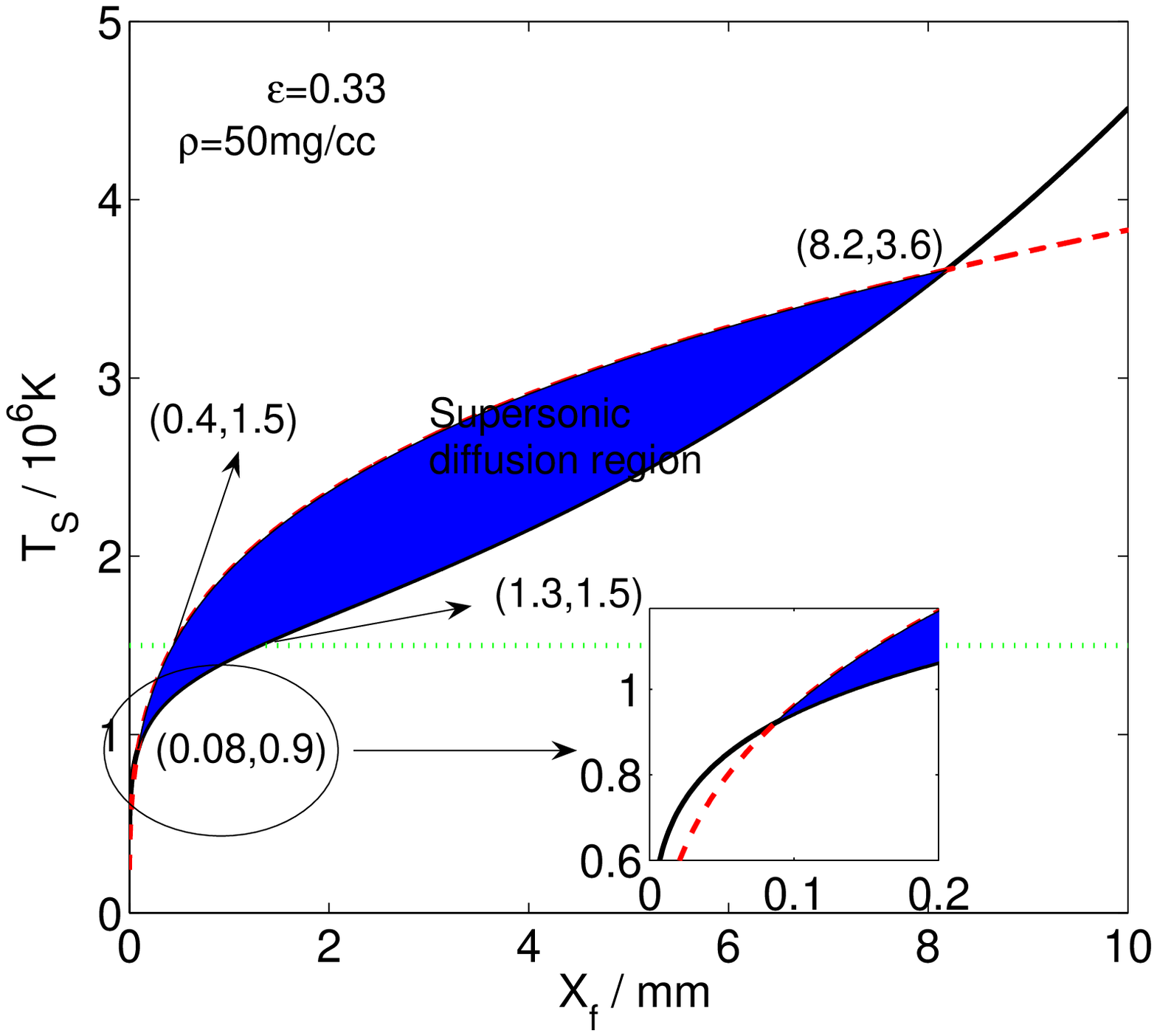,width=9cm,height=7cm}
\caption{Curves for Mach number condition and the optical depth condition with the $\rm SiO_2$ density 50mg/cc and the parameter $\ep=0.29$, the solid line corresponds to the supersonic condition, the dash line corresponds to the optical depth condition and the dotted line represents temperature $T=1.5\times10^6k$.}
\label{Ts1}
\end{figure}

If we fix the density to 50mg/cc, and change the sign $'>'$ in Eq. (\ref{cnsio3}) to $'='$, every source temperature $T_S$ corresponds to a front position $x_{1}$ according to the first condition, and according to the second condition, corresponds to another front position $x_{2}$.
These temperature and front position are plotted in FIG. \ref{Ts1}. the solid line corresponds to the supersonic condition, the dash line corresponds to the optical depth condition and the dotted line represents temperature $T=1.5\times10^6k$. The condition $M>\frac{8}{3}(1+\ep/3)$ corresponds to the the region above the solid line, and the condition $\tau>1$ corresponds to the region below the the dash line. Noticing the FIG. \ref{Ts1}, when setting the parameter $\ep=0.33$ which indicates there is $38\%$ of the radiation energy flux absorbed by the wall, the dash line and the solid line cross each other two times, at the point (0.08, 0.93)
and (8.2, 3.6) respectively. As long as the point ($T_s$,$x_F$) locates in the blue region shown in the picture, the supersonic radiation diffusion approximation can be used. When the temperature is as low as 0.9heV, the supersonic region comes into being. When the temperature becomes higher and tends to be 2heV, the distance supersonic heat wave propagate tends to be longest. When the temperature is around 3.6heV, we can use supersonic radiation diffusion no longer. When setting the temperature to be $1.5\times10^6k$, the supersonic radiation diffusion region is $0.44mm<x<1.3mm$. Therefore, it is reasonable to study the supersonic radiation diffusion with the source temperature $\sim1.5heV$ and the length of sample $~0.5mm-1.25mm$.

\begin{figure}
\psfig{figure=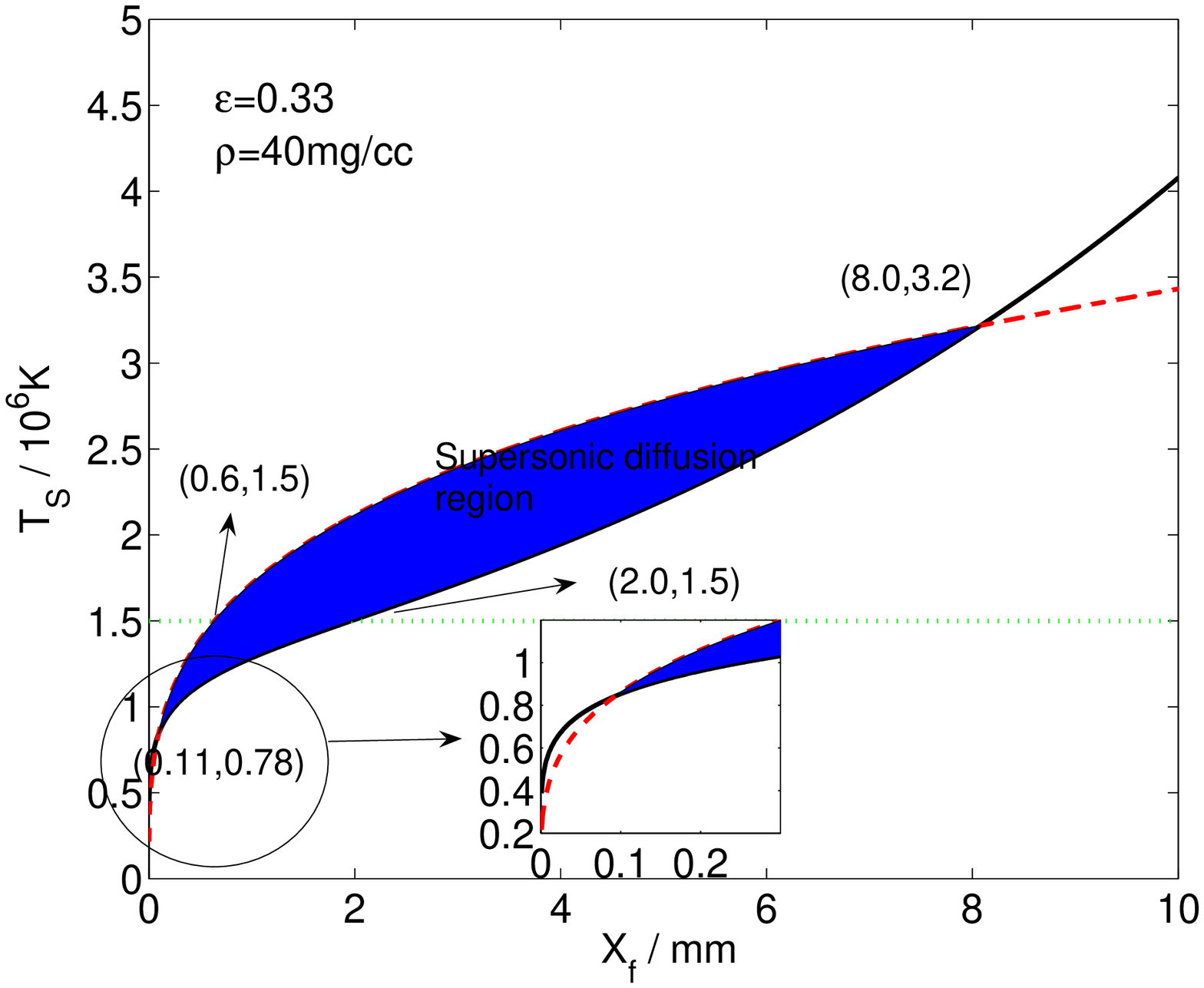,width=7cm,height=7cm}\psfig{figure=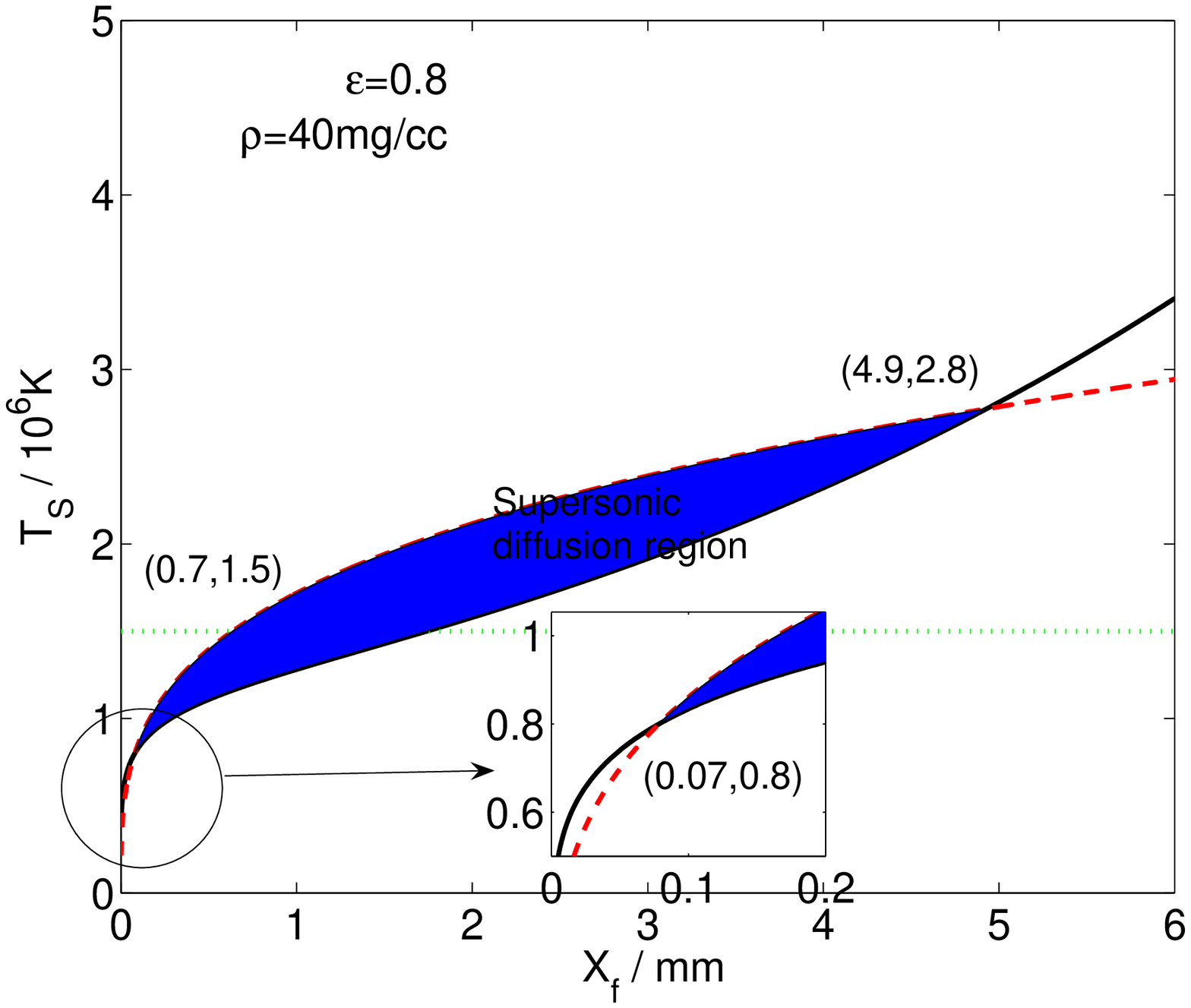,width=7cm,height=7cm}
\caption{Curves for Mach number condition and the optical depth condition with the $\rm SiO_2$ density 40mg/cc, with the parameter $\ep=0.33$(left) and $\ep=0.8$(right), the solid line corresponds to the supersonic condition, the dash line corresponds to the optical depth condition and the dotted line represents temperature $T=1.5\times10^6k$.}
\label{Ts2}
\end{figure}

In the same way, when fixing the density to 40mg/cc and also set $\ep$ to be 0.33, we get a similar supersonic region, seeing FIG. \ref{Ts2}(left). However the total region has a shift in the left-down direction. When setting the temperature to be $1.5\times10^6k$, the supersonic radiation diffusion region is $0.6mm<x<2.0mm$.

When fix the density to 40mg/cc and also set $\ep$ to be 0.8, we get a narrower supersonic region comparing to the case of $\ep=0.33$, seeing FIG. \ref{Ts2}(right). The upper limit of temperature becomes smaller while the lower limit keep the same with that in case of $\ep=0.33$. It indicates that a bigger $\ep$ which means more energy is absorbed by walls makes the supersonic region narrower.

\subsection{supersonic diffusion region for ${\rm Au}$}
\begin{figure}
\psfig{figure=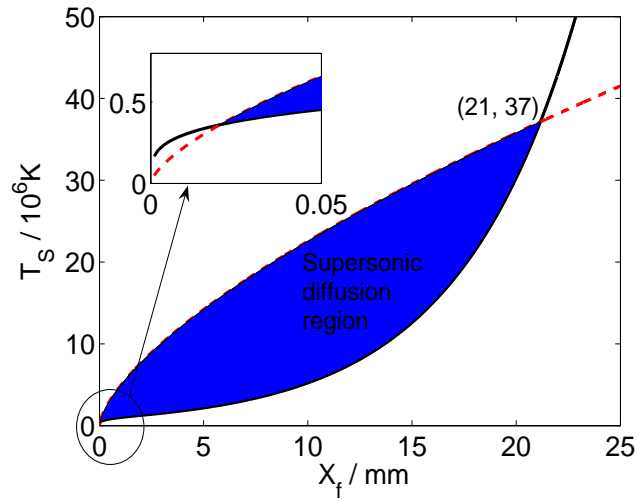,width=9cm,height=7cm}
\caption{Curves for Mach number condition and the optical depth condition with the Au density 50mg/cc and parameter $\ep=0.33$.}
\label{TsAu1}
\end{figure}

Taking Au for example, and knowing the average degree of ionization is about 50 at the temperature 1.9heV\cite{Zhang}, the sound speed can be given as
\bea
\label{Cs1}
Cs(T)\simeq3.36\times10^4[m/s]\left(\frac{T}{10^6k}\right)^{1/2}.
\eea
According to the result of Hammer\cite{Hammer}, the opacity and the specific internal energy can be described as
\bea
\begin{aligned}
\kR&=395.5m^2/kg\left(\frac{T}{10^6k}\right)^{-1.5}\left(\frac{\rho}{50mg/cc}\right)^{0.2}\\
e&=2235[MJ/kg]\left(\frac{T}{10^6k}\right)^{1.6}\left(\frac{\rho}{50mg/cc}\right)^{-0.14}.
\end{aligned}
\label{kRAu}
\eea
In the same way, substituting these expressions into the supersonic conditions, we get
\bea
\label{cnsio2}
\begin{aligned}
1.28\left(\frac{T_s}{10^6k}\right)^{3.9}\left(\frac{\rho}{50mg/cc}\right)^{-2.06}\frac{\sqrt{\ep}}{\sinh(\sqrt{\ep}x_f/L)}>\frac{8}{3},\\
19.75\frac{x_f}{mm}\left(\frac{T_s}{10^6k}\right)^{-1.5}\left(\frac{\rho}{50mg/cc}\right)^{1.2}>1
\end{aligned}
\eea
which means Au has a higher opacity with the same temperature and density comparing to $\rm SiO_2$. According to these two conditions, setting the parameter $\ep=0.33$, and $\rho=50mg/cc$ we give the supersonic diffusion region, seeing FIG. \ref{TsAu1}. These two lines also cross two times, while the supersonic region is much larger than $\rm SiO_2$ thanks to the larger opacity. Noticing the picture, we can see, when the temperature is as low as 0.4heV, the supersonic region comes into being. When the temperature becomes higher and tends to be 10heV, the distance supersonic heat wave propagate tends to be longest. When the temperature is around 37heV, we can use supersonic radiation diffusion no longer .

\section{Conclusions and discussion}
\label{cad}
In the supersonic radiation diffusion theory, the radiant energy flux dominates the material energy flux that gives a limit about the Mach number and the optical depth. By using the 1-dimensions theory, Back et al give this limit as $M\tau>3$\cite{Back} and point out that the density perturbation is less than $10\%$ when Mach number $M>3$. The conditions for supersonic and the optical depth corresponds to a region about the source temperature $T_s$ and the location of wave front $x_f$. Comparing to Back's result, considering the bent Marshak wave theory in two dimensions, we get the supersonic radiation diffusion conditions which are about the Mach number $M>8(1+\sqrt{\ep})/3$, and the optical depth $\tau>1$. It indicates the Mach number depend on how much energy are lost at the wall $y=\pm L$. Obviously, when more radiation energy are lost at the wall, a lager Mach number is required to ensure the radiation energy flux dominates. It is worth mentioning that, for a given density, the theory in 2-dimension gives both lower and upper limit about source temperature and the length of the sample, which makes sure the supersonic heat wave exists, while the theory in 1-dimension only gives a lower limit. This meas the theory in 2-dimension makes the conditions for supersonic diffusion wave stricter comparing to that in 1-dimension.

\section*{Acknowledgments}
This work was in part supported by the NSFC Grant No. 11135007.

\end{document}